\documentclass[12pt]{iopart}

\usepackage{graphics}
\usepackage{flushend}
\usepackage{cuted}
\usepackage{bm}

\usepackage{subfigure}
\usepackage[dvips]{graphicx}
\usepackage{color}
\usepackage{sidecap}

\usepackage{graphicx}
\usepackage{epstopdf}
\usepackage{bm}
\usepackage{subfigure}
\usepackage{sidecap}

\newcommand{\si}{\sigma}
\newcommand{\al}{\alpha}
\newcommand{\bt}{\beta}
\newcommand{\De}{\Delta}

\newcommand{\vare}{\varepsilon}

\newcommand{\be}{\begin{equation}}
\newcommand{\ee}{\end{equation}}
\newcommand{\bea}{\begin{eqnarray}}
\newcommand{\eea}{\end{eqnarray}}
\newcommand{\non}{\nonumber}

\newcommand{\hh}{\hat{h}}

\newcommand{\hG}{\hat{G}}
\newcommand{\hV}{\hat{V}}

\newcommand{\hbx}{\hat{{\bf x}}}

\newcommand{\ua}{\uparrow}
\newcommand{\da}{\downarrow}
\def\dg{^{\dagger}}

\newcommand{\tilt}{\tilde{t}}
\newcommand{\tL}{\tilde{\Lambda}}

\newcommand{\tbq}{\tilde{{\bf q}}}
\newcommand{\tq}{\tilde{q}}
\newcommand{\om}{i\omega_n}

\newcommand{\CC}{CeCoIn$_5$}

\def\ket#1{\left\vert #1 \right\rangle}

\def\br{{\bf r}}
\def\bk{{\bf k}} \def\bq{{\bf q}} 
 \def\hg{\rm\hat{g}} 
\def\bQ{{\bf Q}}

\def\hbx{\hat{{\bf x}}}  

\def\da{\downarrow} \def\ua{\uparrow}
 
\def\dg{\dagger}

\begin{document}

\title[Momentum space imaging of the FFLO state]{Momentum space imaging of the FFLO state}

\author{Alireza Akbari$^{1,2}$ and  Peter Thalmeier$^{3}$}

\address{
$^1$Asia Pacific Center for Theoretical Physics, Pohang, Gyeongbuk 790-784, Korea\\
$^2$Department of Physics, and Max Planck POSTECH Center for Complex Phase Materials, POSTECH,
Pohang 790-784, Korea\\
$^3$Max Planck Institute for the  Chemical Physics of Solids, D-01187 Dresden, Germany
}

\ead{alireza@apctp.org}

\begin{abstract}
In a magnetic field superconductors (SC) with small orbital effect exhibit the Fulde-Ferrell-Larkin-Ovchinnikov (FFLO) phase above the Pauli limiting field. It is characterized by  Cooper pairs with finite center of mass momentum and is stabilized by the gain in Zeeman energy of depaired electrons in the imbalanced Fermi gas. The ground state is a coherent superposition of paired and depaired states. This concept, although central to the FFLO state lacks a direct experimental confirmation. We propose that STM quasiparticle interference (QPI) can give a direct momentum space image of the depaired states in the FFLO wave function.  For a proof of principle we investigate a 2D single orbital tight binding model with a SC s-wave order parameter. Using the equilibrium values of pair momentum and SC gap we calculate the spectral function of quasiparticles and associated QPI spectrum as function of magnetic field. We show that the characteristic depaired Fermi surface parts appear as a fingerprint in the QPI spectrum of the FFLO phase and we demonstrate its evolution with field strength. Its observation in STM experiments would constitute a direct proof for FFLO ground state wave function.
 \end{abstract}

\maketitle

\section{Introduction}

A superconductor with  Pauli limiting behaviour, i.e. weak orbital pair breaking, may exhibit a distinct high-field phase where Cooper pairs have a finite center of mass (CM) momentum. These inhomogeneous  FFLO phases \cite{fulde:64,larkin:65} are therefore characterized by a spatially dependent SC order parameter with a wavelength that corresponds to the CM pair momentum $\bQ=2\bq$. This leads to a real space dependence $\langle \psi^\dg_\da(\br) \psi^\dg_\ua(\br)\rangle \sim \Delta(\br) =\De_\bq\exp(i\bQ\br)$   of the SC phase in the FF state. Here we indexed the gap amplitude by half the CM pair momentum for convenience. Such phases are stabilized by the gain in Zeeman energy due to population imbalance of up and down spin electrons which for large fields overcompensates the loss in pairing energy due to finite CM momentum \cite{combescot:07,zwicknagl:11}. The degeneracy of states with $\pm\bQ$ is lifted by forming a superposition leading to the LO state  $ \Delta(\br) =2\De_\bq\cos(\bQ\br)$.  
In contrast to the unconventional BCS order parameter which changes sign at nodal positions in momentum space, the LO gap function changes sign at real space nodes along the direction of \bQ.  Actually more complicated superpositions are possible and stable at high fields \cite{matsuda:07,shimahara:98} .  In the following we generically denote these states as FFLO even though our concrete calculations will use the FF state only for simplicity. 
Because low dimensionality may further stabilize the FFLO state and increase the  region of stability in the H-T phase diagram in 1D \cite{machida:84} and 2D \cite{shimahara:94,vorontsov:05,baarsma:15} we will restrict to the 2D case which is the most natural choice in our context. 

Nevertheless this phase is hard to observe  experimentally  because it is susceptible to normal impurity scattering \cite{matsuda:07,takada:70} although recent theoretical investigations \cite{wang:07,ptok:10} indicate that the FFLO state may survive when the order parameter profile is relaxed to adapt at scattering sites.  Furthermore orbital pair breaking effects, i.e., the Abrikosov vortex state always destabilize the FFLO phase \cite{gruenberg:66,adachi:03}.  Therefore only few  superconductors have been found as possible candidates. Aside from some organic superconductors \cite{lortz:07,mayaffre:14} the unconventional heavy fermion compound \CC~ is considered to be the most promising case for a FFLO state \cite{matsuda:07} but the ground state is more complicated there due to a coexisting spin density wave \cite{kenzelmann:10,gerber:14,yanase:09,mierzejewski:09}.  Furthermore iron based superconductors \cite{terashima:13,burger:13,zocco:13} are considered as candidates for a multi-band version of the FFLO state. The latter was also recently investigated theoretically \cite{ptok:13,mizushima:14,nakamura:15}. Finally we mention that FFLO states are also intensely discussed in the context of condensed atomic gases \cite{sheehy:07,sheehy:15}. 

A large effort has been made for the experimental characterization of this exotic SC state  \cite{matsuda:07}.   Thermodynamic methods like specific heat, penetration depth and magnetostriction measurements were used to map out the phase boundaries, ultrasonic attenuation \cite{watanabe:04}, thermal transport \cite{capan:04}  and NMR methods \cite{kumagai:06} are powerful tools to investigate the quasiparticle properties.  However, these methods are indirect in the sense that they try to probe the average effect of the normal quasiparticle excitations in the FFLO state without momentum resolution. Therefore these methods do not directly identify the microscopic essence of this phase: The coherent superposition of paired and unpaired electron states on the Fermi surface.\\

Recently the method of STM-quasiparticle interference (QPI) has been employed successfully to investigate the \bk-space node structure of unconventional, in particular heavy fermion superconductors \cite{akbari:11,allan:13,zhou:13}. In this method the scattering of quasiparticles from surface impurities leads to voltage $eV=\omega$ dependent ripples in the local density of state $\delta N_c(\br,\omega)$ of conduction electrons reminiscent of Friedel oscillations. Their momentum Fourier transform, which is the QPI spectrum contains information on Fermi surface (FS) and gap structure \cite{mcelroy:03,capriotti:03}. Most importantly at the nodal positions quasiparticle pockets appear at finite bias voltage that leave a typical signature in this spectrum that can be used to pinpoint the node structure of the gap.

In this work we propose to use this method as a momentum resolved microscopic tool  to investigate the pairing structure of the  FFLO phase. Instead of having just node positions one has finite segments (in 2D) where the pairing amplitude vanishes. This should also leave a strong signature in the QPI structure from which the location and size of unpaired segments of the FS as well as the orientation of the CM momentum can be deduced. This would amount to a complete momentum resolved determination of the FFLO order parameter and not only indirect evidence from the presence of normal quasiparticles.
For a proof of principle of this method we use the simplest type of 2D model, a one orbital tight binding band with an isotropic gap function. This allows to understand the expected fingerprint of the FFLO state in the QPI spectrum without further complications. 
 We note that with a different STM method using directly Josephson or pair-tunneling current recently an inhomogenous superconducting state \cite{hamidian:15} was observed. However, it is not of the FFLO type but is induced at zero field as secondary effect of a primary charge density wave. 
 
In Sect. \ref{sec:model} we define the model which will be diagonalized in Sect.~\ref{sec:bogol}. Using the FFLO Green's function derived in Sect. \ref{sec:green} the QPI spectrum will be calcualted in t-matrix approximation in Sect.~\ref{sec:qpi}. The discussion and interpretation of numerical results is given in Sect.~\ref{sec:discussion} and the conclusion and outlook  in Sect.~\ref{sec:outlook}.

\section{Extended BCS model for finite momentum pairing}
\label{sec:model}

In the canonical BCS state electrons in states $\ket {\bk\ua}$,  $\ket {-\bk\da}$ are bound in spin singlet Cooper pairs with vanishing center of mass momentum $\bQ=0$ which has the lowest energy in zero field. In an imbalanced Fermi gas where the number of up and down spin electrons is unequal due to the application of an effective field (either external or molecular) the FFLO pair state with finite CM momentum $\bQ=2\bq$ may become more stable provided the orbital effect of the field is negligible \cite{fulde:64,larkin:65}. Such a ground state consists of a coherent superposition of paired and unpaired electrons according to \cite{sheehy:07,sheehy:15,cui:06} 
\bea
\ket {\Psi_\bq}=
\prod_{-\bk \in \bk_3}c^\dag_{-\bk+\bq\da}
\prod_{\bk \in \bk_2}c^\dag_{\bk+\bq\ua}
\prod_{\bk \in \bk_1}\Bigl(u_\bk+v_\bk c^\dag_{\bk+\bq\ua} c^\dag_{-\bk+\bq\da} \Bigr)\ket{0},
\label{eq:gswave}
\eea
where 1 refers to the momenta with finite BCS pairing amplitude   $v_\bk\equiv v_{\bk\bq}$   and 2,3 to momenta with unpaired electrons. 
Their relative size determines the stability of the FFLO ground state at a given field with respect to the BCS state which has no unpaired regions 2,3. They may be determined by the diagonalization of the generalized ($\bq\neq 0$) SC mean field Hamiltonian.  As explained in Sec.~\ref{sec:green} paired region 1 is characterized by positive and unpaired regions 2,3 by negative Bogliubov energies of the diagonalized Hamiltonian. 
The concept of paired and unpaired regions of the Fermi surface is fundamental to the FFLO state \cite{fulde:64}, however it has sofar not been identified directly in an experiment. Here we propose a way to study the momentum space image of the FFLO state using STM-QPI method.\\

It is known that low dimensionality (D=1,2) helps to stabilize the FFLO superconducting state \cite{machida:84,shimahara:94,vorontsov:05}. Therefore we consider a 2D model which in any case is the most suitable one from the viewpoint of QPI investigation. The mean field Hamiltonian of the 2D superconductor is given by
\bea
H_{MF}=\sum_{\bk\si}\xi^\si_{\bk}c^\dg_{\bk\si}c_{\bk\si}
-\sum_{\bk}(\Delta_{\bk\bq}c^\dg_{\bk+\bq\ua}c^\dg_{-\bk+\bq\da}
+\Delta^*_{\bk\bq}c_{-\bk+\bq\da}c_{\bk+\bq\ua}),
\label{eq:hmf}
\eea
with the Zeeman-split band energies ($\vare_\bk$ = conduction band dispersion, $\mu$=chemical potential) defined by  
\bea
\xi_\bk^\si=\xi_\bk+\si h = \vare_\bk-\mu+\si h\;\; ( \sigma =\pm 1),
\eea
where $h=\mu_BH$ and $H$ is the magnetic field. Furthermore a simple one-band tight binding (TB) dispersion $\vare_\bk =-2t(\cos k_x +\cos k_y) $ with hopping element $t$ and band width $W=2D=8t$ is used. Here we consider only a FFLO state based on s-wave  gap function, i.e., we assume $\De_{\bk\bq}=\De_\bq$. The SC gap function $\De_\bq$ is physically confined inside a cutoff shell $(\xi_c\ll W)$ according to  $\De_{\bq}=\De_{\bq}\Theta(\xi_c-|\xi_{\bk+\bq}|)\Theta(\xi_c-|\xi_{\bk-\bq}|)$. We will later absorb the cutoff into the interaction constant $V_0$ for ease of numerical computation. The total extended SC mean field Hamiltonian for finite CM momentum $\bQ=2\bq$ then reads
\bea
H_{SC}=
H_{MF}+N\Bigl(\frac{|\De_\bq|^2}{V_0}\Bigr)=
\sum_{\bk}\psi^\dg_{\bk\bq}\hh_{\bk\bq}\psi_{\bk\bq}
+E_0 + N\Bigl(\frac{|\De_\bq|^2}{V_0}\Bigr),
\non\\
\hh_{\bk\bq}=
\left(
 \begin{array}{cc}
 \xi_{\bk+\bq\ua}&-\Delta_{\bq} \\
 -\Delta_{\bq}^*& -\xi_{-\bk+\bq\da}
\end{array}\right),
\non\\
 E_0=\sum_{\bk}\xi^\da_{\bk+\bq},
\eea
where we used the representation with Nambu spinors defined by $\psi^\dg_{\bk\bq}=(c^\dg_{\bk +\bq\ua},c_{-\bk +\bq\da})$.
%
\begin{figure}
\centerline{
\includegraphics[width=0.95\linewidth]{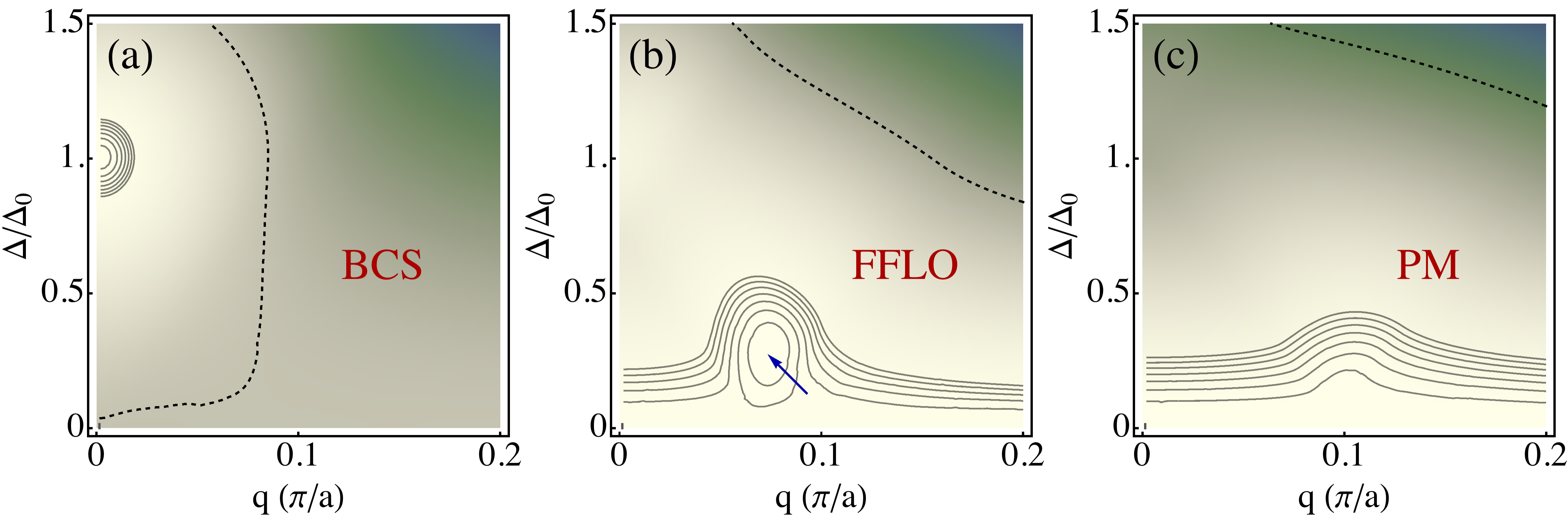}
}
\caption{
Topography of condensation energy $E_c$ (with respect to $h=0,\bq=0$ state) in $(q,\Delta_q)$-plane (here $\Delta_q$ in 
units of $\Delta_0=0.5t$ and  $\bq={q}\hbx$, with $q$ labeled in units of $\pi/a$). Light (dark) color correspond to small (large) 
values of $E_c$. For absolute minima in (a) and (b) $E_c<0$.  (a) $h=0$: absolute minimum at BCS state ($q=0, \De_q=\De_0$)  
(b) $h=0.8\Delta_0$: absolute minimum at FFLO state ($q= 0.08 ,\De_q= 0.2\De_0$) in center of FFLO region,  
(c) $h=\De_0$: normal  and (polarized) paramagnetic (PM)  state $(\De_q=0)$.  Here $E_c$ is negative (positive) to the left (right)
of the dotted lines. 
}
\label{fig:Fig1}
\end{figure}
%

\section{Ground state energy and Bogoliubov quasiparticles}
\label{sec:bogol}

The proper CM momentum of the FFLO state is determined by the minimization of the ground state energy which is most conveniently obtained by diagonalizing $H_{SC}$ with a Bogoliubov transformation \cite{sheehy:07,cui:06}. For the paired states $\bk\in\bk_1$ it is defined by 
\bea
\left(
\begin{array}{c}
c_{\bk+\bq\ua}\\
c^\dg_{-\bk+\bq\da}
\end{array}\right)
&=&
\left(
 \begin{array}{cc}
 u^*_\bk&v_\bk \\
 -v^*_\bk& u_\bk
\end{array}\right)
\left(
\begin{array}{c}
\alpha_\bk\\
\beta^\dg_\bk
\end{array}\right),
\label{eq:bogol1}
\eea
where $\alpha_\bk,\beta_\bk$ are the quasiparticle operators in the FFLO state. The transformation matrix is given by 
\bea
|u_\bk|^2=\frac{1}{2}\Bigl(1+\frac{\xi^s_{\bk\bq}}{E_{\bk\bq}}\Bigr)=1-|v_\bk|^2;\;\;\;
E_{\bk\bq}=\sqrt{\xi^{s2}_{\bk\bq}+|\De_\bq|^2},
\label{eq:bogol2}
\eea
where we defined (anti-) symmetrized kinetic energies  with respect to half of the pair momentum \bq:
\bea
\xi^{s}_{\bk\bq}&=&\frac{1}{2}(\xi_{\bk+\bq} + \xi_{\bk-\bq});\;\;\;
\xi^{a}_{\bk\bq}=\frac{1}{2}(\xi_{\bk+\bq} - \xi_{\bk-\bq})
.
\eea
Using similar transformations  for the unpaired regions $\bk\in\bk_2,\bk_3$ the Hamiltonian may be  diagonalized \cite{sheehy:07,cui:06}  to give
\bea
H_{SC}&=&E_G(\bq,\De_\bq)+\frac{1}{2}\sum_{\bk}(|E^+_{\bk\bq}|\al^\dg_\bk\al_\bk+|E^-_{\bk\bq}|\bt^\dg_\bk\bt_\bk)
,
\label{eq:BCS2}
\eea
where the first part is the ground state energy and the second part describes the quasiparticle excitations for $\bk$ in any of the three regions.  The (positive) excitation energies of the latter are given by  $|E^\si_{\bk\bq}|$ with $(\sigma=\pm)$:
\bea
E^\si_{\bk\bq}=E_{\bk\bq}+\si(\xi_{\bk\bq}^a+h)
.
\label{eq:quasienergy}
\eea
In the paired $\bk_1$ momentum space region  $E^\si_{\bk\bq}>0$. In the unpaired $\bk_2$ region $E^-_{\bk\bq}<0$ and $E^+_{\bk\bq}<0$ in the unpaired $\bk_3$ region. Note that although in regions 2,3 electrons are unpaired their quasiparticle energy nevertheless contains the gap function determined solely by the electrons in region 1. This is a consequence of the coherent superposition of paired and unpaired states in the FFLO ground state wave function in Eq.~(\ref{eq:gswave}).   For the gap size  $\Delta_\bq$ we only need to consider the zero temperature case which is relevant for QPI. The corresponding s-wave gap equation for $\Delta_\bq$ (isotropic in \bk) is \cite{shimahara:98}
\bea
\Bigl(\frac{1}{V_0}\Bigr)=\frac{1}{N}\sum_\bk \frac{1}{2E_{\bk\bq}}\theta(E^+_{\bk\bq})\theta(E^-_{\bk\bq})
,
\label{eq:gapfflo}
\eea
%
%
\begin{SCfigure}
{\centering
\includegraphics[width=0.50\linewidth]{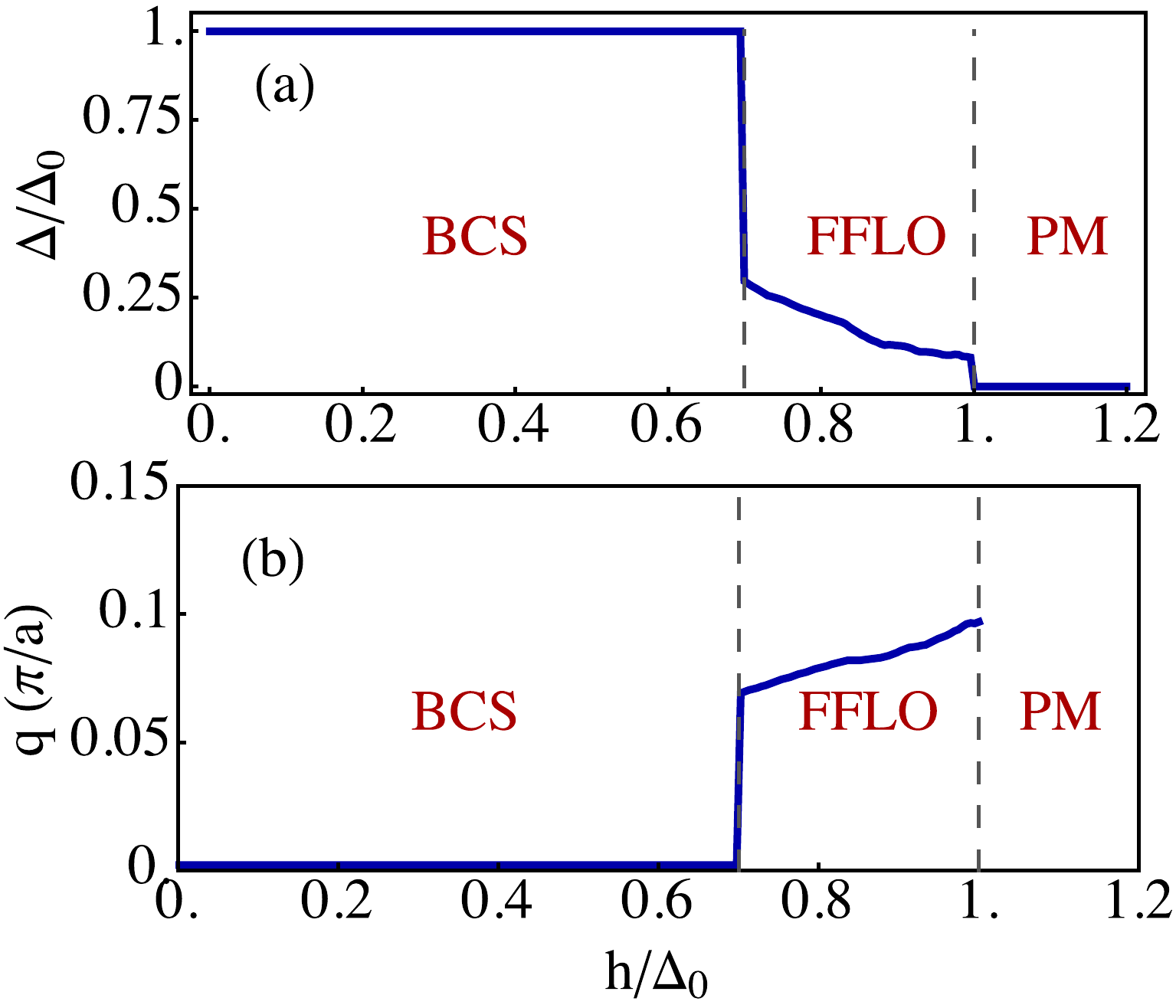}
\caption{Dependence of gap size $\Delta_q$ (a) and half CM pair momentum (b) on the magnetic field. FFLO phase sets in for $h>h_p=0.7\De_0$ 
and is destroyed at $h_c=\Delta_0$. A tight binding band model with band width $W=8t$ and chemical potential $\mu=-2.8t$ has been used. BCS 
gap size was choosen $\De_0=0.5t$.
\vspace{1cm}
}
}
\label{fig:Fig2}
\end{SCfigure}
%
where we used the $T\rightarrow 0$ equivalence $1-\sum_\sigma f(E^\sigma_{\bk\bq})=\theta(E^+_{\bk\bq})\theta(E^-_{\bk\bq})$ for the Fermi function $f(E)$.
Here the summation is restricted to the paired momentum region with both $E^\si_{\bk\bq}>0$ due to the Heaviside functions and $\Delta_\bq$ appears implicitly in $E_{\bk\bq}$ (Eqs.~(\ref{eq:bogol2},\ref{eq:quasienergy})).
The ground state energy in Eq.~(\ref{eq:BCS2}) finally is given in the form \cite{sheehy:07}
\bea
\hspace{-1.5cm}
E_G(\bq,\De_\bq)=N\Bigl(\frac{|\De_\bq|^2}{V_0}\Bigr)-\sum_\bk(E_{\bk\bq}-\xi^s_\bk)
+\sum_\bk[E^+_{\bk\bq}\theta(-E^+_{\bk\bq}) +   E^-_{\bk\bq}\theta(-E^-_{\bk\bq})]    
.
\label{eq:gsen}
\eea
This energy functional has to be minimized with respect to $\bq$ and $\De_\bq$ to find the true ground state at a given field strength h. It comprises the FFLO state $(|\bq|>0,|\De_\bq|>0)$ , the  BCS state $(\bq=0,|\De_\bq|>0)$ and the unpolarized (h=0) normal state  $(\bq=0,|\De_\bq|=0)$. In the latter case it simplifies to 
\bea
E^0_G=E_G(0,0)=\sum_\bk(\xi_\bk-|\xi_\bk|)=\sum_{\bk\si}f_{\bk}\xi_\bk
,
\label{eq:gsen0}
\eea
where $f_\bk=\Theta(-\xi_\bk)$ is the zero temperature Fermi function.
Subtracting this from Eq.~(\ref{eq:gsen}) and rearranging terms we obtain the condensation energy $E_c=E_G-E^0_G $ (refered to the $h=0, \bq =0 $ normal state energy $E_G^0$) as 
\bea
E_c(\bq,\De_\bq)&=&N\Bigl(\frac{|\De_\bq|^2}{V_0}\Bigr)
-\sum_\bk(E_{\bk\bq}-|\xi_\bk|)+\sum_\bk(\xi^s_{\bk\bq}-\xi_\bk)
\nonumber
\\&&
+\sum_\bk[E^+_{\bk\bq}\theta(-E^+_{\bk\bq}) +   E^-_{\bk\bq}\theta(-E^-_{\bk\bq})]    
.
\label{eq:gscond}
\eea
This form is most suitable for the numerical calculations. The ground state as a function of field h will be determined by directly minimizing the condensation energy (i.e. maximizing its absolute value). For that purpose it is more convenient to eliminate the pairing strength $V_0$ by assuming a fixed $h,\bq=0$ BCS gap amplitude $\Delta_0$ using the gap equation for that case: 
\bea
\frac{1}{V_0}=\frac{1}{N}\sum_{\bk}\frac{1}{2E_\bk}\Theta(\xi_c-|\xi_\bk|)
,
\label{eq:gapbcs}
\eea
where now $E_\bk=\sqrt{\xi^2_\bk+\De_0^2}$. Here $\xi_c$ is a Debye cutoff $(\De_0<\xi_c<W)$. Since we keep a fixed $\Delta_0$ we can absorb the cutoff into a renormalized coupling constant by dropping $\Theta$.\\

%
\begin{SCfigure}
\includegraphics[width=0.50\linewidth]{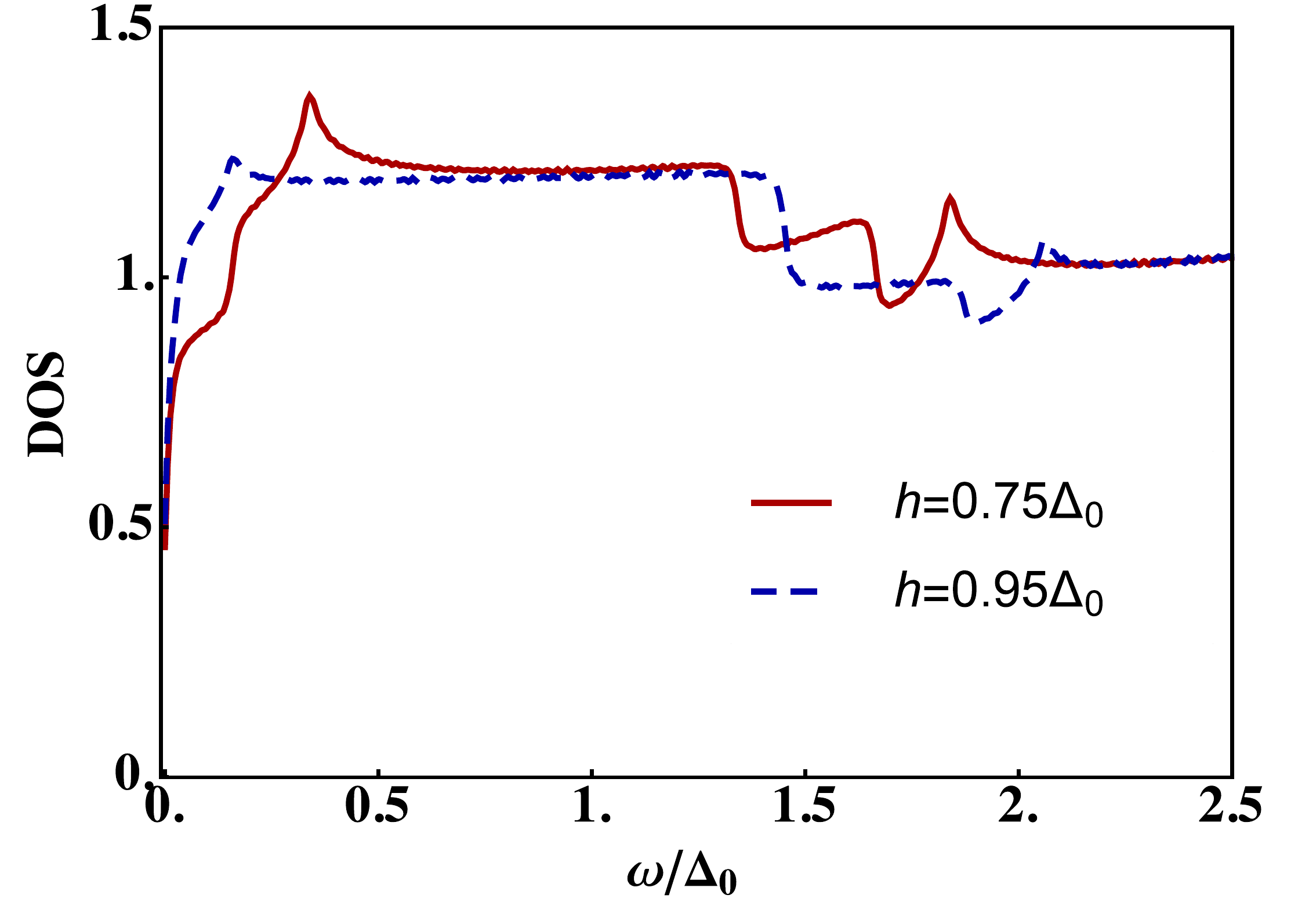}
\caption{
Quasiparticle DOS in the FFLO phase at different fields. In the low energy part the FFLO gap $\Delta_q$ which decreases with increasing $h$ is visible.
 However a large zero energy  DOS $\rho(\omega=0)$ remains due to the depaired states. At higher energy a remnant anomaly due to the  (local) 
 BCS gap $\De_0$ occurs.
 \vspace{0.5cm}
 }
\label{fig:Dos_75-95}
\end{SCfigure}
%
Now  we discuss the field dependent SC ground state of the model.
For the numerical calculations we choose a tight binding dispersion to assure lattice periodicity of the final QPI spectrum and to avoid introducing an artificial energy cutoff in the ground state energy calculation. The overall energy scale is the band width $W=8t$ with hopping energy $t\equiv 1$. 
The chemical potential $\mu=-2.8t$ is chosen to obtain a small, almost spherical FS for ease of interpreting QPI. In this case we can choose $\bq=q\hbx$ without restriction.
For the maximum BCS gap we use a large value $\Delta_0=0.5t$ to facilitate the discussion of FFLO features in the final QPI spectrum.
The condensation energy $E_c(q,\De_q)$ is shown in three contour plots in the $(q,\De_q)$ - plane in Fig.~\ref{fig:Fig1}. 
 For fields up to $h_p=0.7\De_0$ the absolute minimum is stable at $q=0,\Delta_q=\De_0$, i.e. the BCS state is realized (a). Here $h_p$ is essentially the Chandrasekhar-Clogston or Pauli limiting field $\De_0/\sqrt{2}$ for BCS pair breaking. However, already above 
 $h=0.6\De_0$ a second local minimum develops at finite q and reduced gap, it becomes the absolute minimum for  $h>h_p$ indicating the appearance of a stable FFLO phase above that field (b). Above $h_c =\Delta_0$ the gap closes and the normal state appears (c).
 The variation of absolute minimum position $(q(h),\Delta_q(h))$ with field strength at zero temperature is extracted from this energy landscape and shown in Fig.~\ref{fig:Fig2} for the 2D TB model. It demonstrates the appearance of the FFLO phase above $h_p$ and the gap closing to the normal state above $h_c$.
  The transition at $h_p$ is of first order \cite{combescot:07,shimahara:94} and the transition to the normal state at $h_c$ should be of second order  with diverging slope of $\De_\bq$ \cite{machida:84}. Due to the limited accuracy of the numerical 2D integration we cannot really resolve this behaviour. 
The zero temperature limit considered here is in fact singular at $h_c$ and the true ground state was proposed as a cascade of FFLO type states with ever more increasing higher harmonic content \cite{mora:04}.

\section{The FFLO Green's function}
\label{sec:green}

For calculation of the momentum dependent QPI spectrum we need the Green's function in the FFLO state. Using inversion symmetry $\xi_{-\bk+\bq} = \xi_{\bk-\bq} $ it is given by
\bea
\hspace{-1cm}
\hG_{\bq}(\bk,\om)=
(\om-\hh_{\bk\bq})^{-1}=
\frac{1}{D_{\bk\bq}(\om)}
\left(
 \begin{array}{cc}
 \om+\xi^\da_{\bk-\bq}   &-\Delta_{\bq} \\
 -\Delta_{\bq}^*& \om-\xi^\ua_{\bk+\bq}
\end{array}
\right).
\label{eq:greenmat}
\eea
%
%
\begin{figure}
\centerline{
\includegraphics[width=\linewidth]{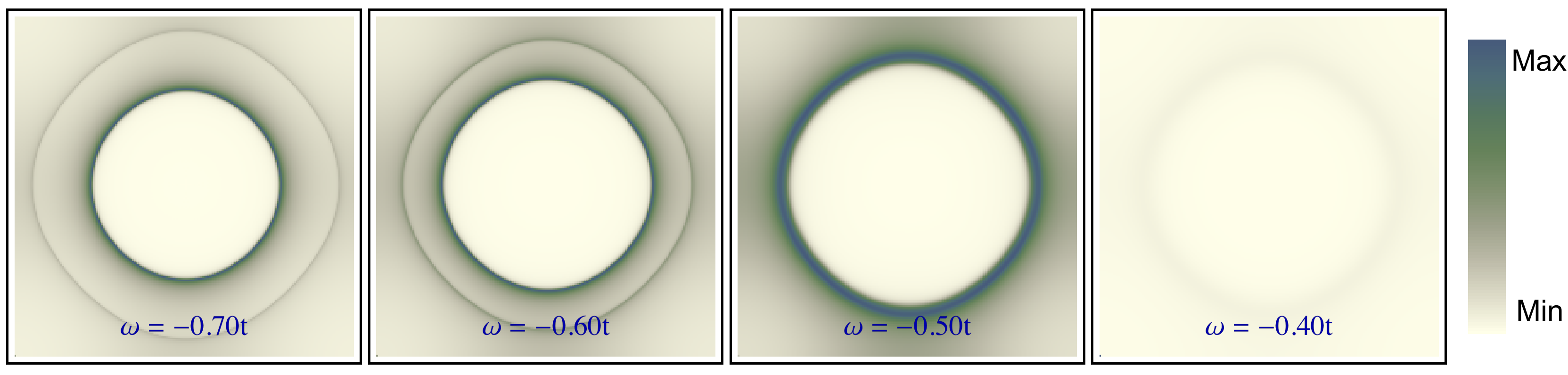}
}
\caption{QPI spectrum for simple  BCS case $(h,\bq=0)$  with $E_\bk=\sqrt{\xi^2_\bk+\De_0^2}$ in $(\tq_x,\tq_y)$-plane (units: $\pi/a$). 
Here $\De_0=0.5t$. Double circles are due to square-root behaviour of quasiparticle energy  $E_\bk$. Therefore they coalesce at $\omega  = -\De_0$ and become extinct for $|\omega| < \De_0$. The QPI picture for $\omega > 0$ is similar. Here and in the following figures we plot the absolute value of QPI spectrum. }
\label{fig:QPI_BCS}
\end{figure}
%
The denominator in  Eq.~(\ref{eq:greenmat})  may  be written in different forms:
\bea
D_{\bk\bq}(\om)&=&(\om-\xi^\ua_{\bk+\bq})(\om+\xi^\da_{\bk-\bq})-|\De_\bq|^2\non\\
&=&(\om)^2-\om(\xi^\ua_{\bk+\bq}-\xi^\da_{\bk-\bq})-E^+_{\bk\bq} E^-_{\bk\bq},
\eea
where Eq.~(\ref{eq:quasienergy}) and the identities 
\bea
\xi^s_{\bk\bq}=\frac{1}{2}(\xi^\ua_{\bk+\bq}+\xi^\da_{\bk-\bq}),
\non\\
E_{\bk\bq}=\frac{1}{2}(E^+_{\bk\bq}+E^-_{\bk\bq}),
\non\\
\xi^a_{\bk\bq} +h=\frac{1}{2}(\xi^\ua_{\bk+\bq}-\xi^\da_{\bk-\bq})=\frac{1}{2}(E^+_{\bk\bq}-E^-_{\bk\bq}),
\non
\eea
have been used. Finally  it can be expressed as 
\bea
D_{\bk\bq}(\om)=(\om-E^+_{\bk\bq})(\om+E^-_{\bk\bq}),
\label{eq:Det2}
\eea
which reduces to the usual form $D_{\bk\bq}(\om)=[(\om)^2-E_\bk^2]$ for the conventional BCS state when $h,\bq=0$.
Furthermore the FFLO spectral function of the Green's function in Eq.~(\ref{eq:greenmat}) is given by
\bea
A_{\bk\bq}(\omega)=-\frac{1}{\pi}
{\rm Im}~tr[\hG_{\bq}(\bk,\omega+i\eta)]_{\eta\rightarrow 0} = \delta(\omega -E^+_{\bk\bq})+\delta(\omega +E^-_{\bk\bq}).
\label{eq:DOS}
\eea
In the BCS case $E_{\bk\bq}^\pm=E_{\bk\bq} > 0$. The spectral functions of the model are shown  further below in subsequent  Figs.~\ref{fig:QPI_h075},\ref{fig:QPI_h09} (top panels) at various frequencies (with respect to $\mu$)  for fields $h=0.75\De_0$ and $h= 0.9\De_0$, respectively within the FFLO phase. For a chemical potential $\mu=-2.8t$ the FS (dashed line) is nearly spherical and well inside the Brillouin zone (BZ). According to Eq.~(\ref{eq:gswave}) the FFLO state is a coherent superposition of paired states ($\bk_1$ region)  (Eq.~(\ref{eq:gswave})) with   $E^\pm_{\bk\bq} >0$ and unpaired states ($\bk_2$ region) with   $E^-_{\bk\bq} < 0 $ inside blue line and unpaired states ($\bk_3$ region) with $E^+_{\bk\bq} <0$ inside red line. For the equilibrium values of $(q(h),\De_q(h))$ in the FFLO range of $h_p<h< h_c$ only  $E^-_{\bk\bq} <0$ is possible, i.e. depaired states exist only on the right (blue) sheet \cite{sheehy:15}. This means that  there are always paired states on the left $(k_x < 0)$ that stabilize the FFLO state. 
This refers to the left plots with $\omega =0$. The other figures show how the equal energy contours $\pm E^\pm_{\bk\bq} =\omega$  or spectral function evolves with frequency. The  $-E^-_{\bk\bq} =\omega$ sheet shrinks (grows) for increasing (decreasing) $\omega$. The  $E^+_{\bk\bq} =\omega$ sheet does not exist for $\omega=0$ because $E^+_{\bk\bq}>0$ in the ground state but then appears for positive $\omega$. Since the gap is smaller for larger $h$ in the FFLO phase the red  $E^+_{\bk\bq} =\omega$ sheet  appears more quickly for $h=0.9\De_0$ than for $h=0.75\De_0$ as demonstrated in Figs.~\ref{fig:QPI_h075},\ref{fig:QPI_h09} (top panels).\\

The FFLO quasiparticle density of states (DOS) corresponding to the second term in Eq.~(\ref{eq:BCS2}) is given by \cite{cui:06}
\bea
\rho(\omega>0)=\frac{1}{N}\sum_\bk[ \delta(\omega -|E^+_{\bk\bq}|)+\delta(\omega -|E^-_{\bk\bq}|)].
\eea
In the BCS case $(h=0)$ it reduces to the well known form with a square root singularity above $\De_0$. In the FFLO phase it is plotted for $h=0.75\De_0$ and $h=0.95\De_0$ in Fig.~\ref{fig:Dos_75-95}. It exhibits the presence of the FFLO gap $\De_q$ at low frequency as well as the finite zero energy density of states $\rho(\omega =0)> 0$ due to depaired states.
%
\begin{figure}
\centerline{
\includegraphics[width=0.85\linewidth]{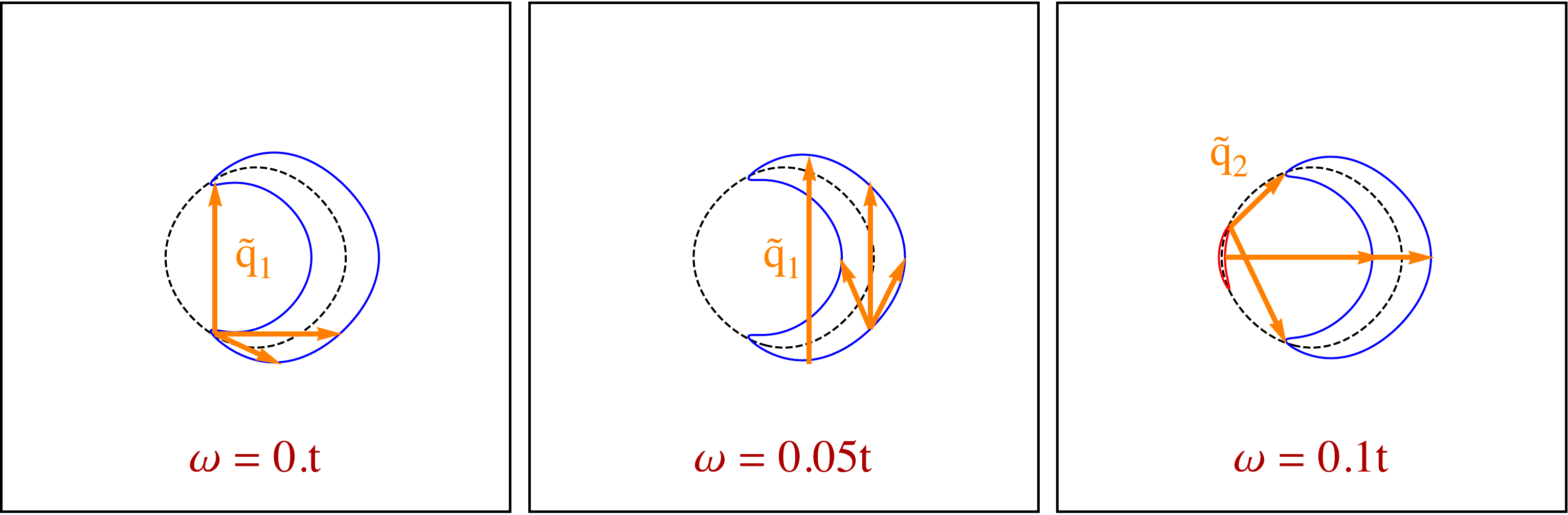}
}
\centerline{
\includegraphics[width=0.85\linewidth]{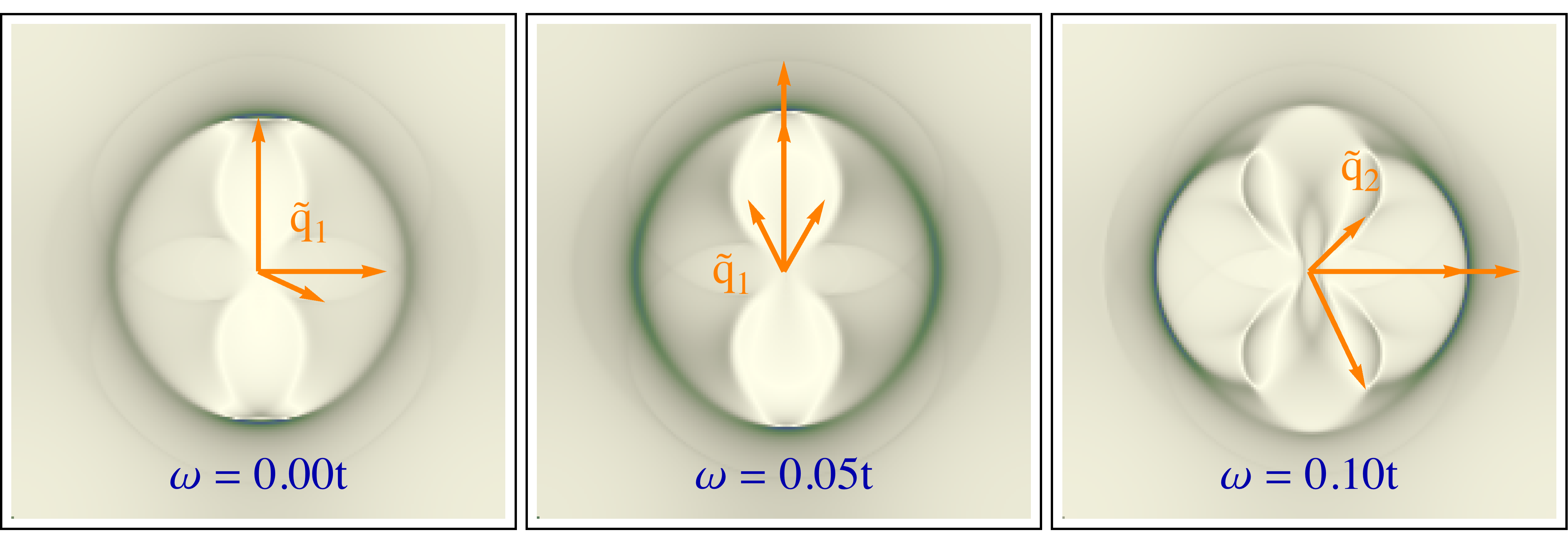}
}
\caption{Spectral function and associated QPI spectrum for $h=0.75\De_0$ with $q(h)= 0.074, \Delta_q(h)=0.243\De_0$.
Top: Spectral function in $(k_x,k_y)$-plane (units: $\pi/a$, range $[-1,1]$). Fermi surface for $\mu=-2.8t$ shown black dashed. We assume $\bq={q}\hbx$. For $\omega=0$ the unpaired regions in \bk-space with $E^-_{\bk\bq} <0$ (blue) correspond to regions inside the full lines. Since $E^+_{\bk\bq} > 0$ unpaired regions inside  $E^+_{\bk\bq} =0$  (red) on the left $(k_x <0)$ do not appear. For finite $\omega$ the evolution of equal energy contours  $\pm E^\pm_{\bk\bq} =\omega$ is shown. For $\omega =0,0.05t$ only intra-sheet (blue) $\tbq_1$= scattering is possible, for
$\omega=0.1t$ both red and blue quasiparticle sheets are present leading also to inter-sheet $\tbq_2$ scattering.
Bottom: Corresponding QPI- spectrum in  $(\tq_x,\tq_y)$-plane (units: $\pi/a$) in Born approximation according to Eq.~(\ref{eq:qpistruc}) for field as above within FFLO phase. Intra- and inter- sheet scattering vectors $\tbq_1$  and $\tbq_2$  are indicated correspondingly.}
\label{fig:QPI_h075}
\end{figure}
%
\section{Calculation of the QPI spectrum}
\label{sec:qpi}

The quasiparticle interference spectrum results from single-site scattering of electrons from surface impurities. The scattered waves interfere with the ingoing waves and cause a pattern of real-space ripples in the spectrally resolved  electron density of states $\delta N_c(\br,\omega)$. Here $\omega=eV$ is the energy set by the bias voltage $V$. Integration over $\omega$ leads to the real-space Friedel oscillations $\delta N_c(\br)$ of the total electron density caused by the impurity scattering. In unconventional superconductors the momentum ($\tbq$) Fourier transform $\delta N_c(\tbq,\omega)$ of the QPI pattern contains important information on the electron dispersion $\vare_\bk$ and the superconducting gap $\De_\bk$ \cite{mcelroy:03,capriotti:03} and in favorable 2D cases with simple FS geometry as in the cuprates allows the determination of both quantities \cite{mcelroy:03}. Here we use a similar approach for the FFLO state with finite CM momentum. In single impurity t-matrix scattering theory the change in the charge density due to surface impurities is obtained as
$\delta N_c(\tbq,\omega)=-(1/\pi){\rm Im} \big[\tL_{0\bq}(\tbq,\omega+i\eta)_{\eta\rightarrow 0}\big]$ with
\bea
\tL_{0\bq}(\tbq,\om)=\frac{1}{2N}\sum_\bk\Bigl[\hG_\bq(\bk,\om)\hat{t}(\om)\hG_\bq(\bk-\tbq,\om)\Bigr]_{11},
\label{eq:QPI-FFLO}
\eea
where (11) denotes the matrix element in Nambu space. Note that here $\tbq$ is the momentum in the 2D QPI Fourier transform whereas $\bQ=2\bq$ is  the fixed CM momentum of the Cooper pairs in the FFLO state at a given field strength h. In addition to the Green's function of Eq.~(\ref{eq:greenmat}) we then need the scattering t-matrix that originate from a (non-magnetic) isotropic impurity scattering potential  $V_c(\tbq)=V_c$  described by  \cite{capriotti:03} $\hV_c=V_c\tau_3$ where $\tau_3$ is a Pauli matrix in Nambu space.
Summing repeated scattering at a single impurity up to infinite order then leads to the $2\times 2$ scattering t-matrix given by 
\bea
\hat{t}_\bq(\om)=
\Big[
1-\hV{\hg_\bq}(\om)
\Big]^{-1}\hV=
\left(
 \begin{array}{cc}
\tilt_d(\om)& t_a(\om) \\
 t_a(\om)& -t_d(\om)
\end{array}
\right),
\label{eq:tmat}
\eea
where $\hg_\bq(\om)=(1/N)\sum_\bk\hG_\bq(\bk,\om)$. The scattering matrix elements are evaluated as
\bea
\tilt_d(\om)&=&\frac{V_c}{d_\bq(\om)}\Bigl[1+\frac{V_c}{N}\sum_\bk\frac{\om-\xi^\ua_{\bk+\bq}}{D_{\bk\bq}(\om)}\Bigr],
\non\\
t_d(\om)&=&\frac{V_c}{d_\bq(\om)}\Bigl[1-\frac{V_c}{N}\sum_\bk\frac{\om+\xi^\da_{\bk-\bq}}{D_{\bk\bq}(\om)}\Bigr],
\non\\
t_a(\om)&=&\frac{V_c}{d_\bq(\om)}\frac{V_c}{N}\sum_\bk\frac{\De_\bq}{D_{\bk\bq}(\om)},
\label{eq:telement}
\eea
where we the determinant in the denominator is given by 
\bea
\hspace{-1.9cm}
d_\bq(\om)=
\Bigl[
1-\frac{V_c}{N}\sum_\bk\frac{\om+\xi^\da_{\bk-\bq}}{D_{\bk\bq}(\om)}
\Bigr]
\Bigl[1+\frac{V_c}{N}\sum_\bk\frac{\om-\xi^\ua_{\bk+\bq}}{D_{\bk\bq}(\om)}
\Bigr]
+
\Bigl[
\frac{V_c}{N}\sum_\bk\frac{\De_\bq}{D_{\bk\bq}(\om)}
\Bigr]^2
.
\eea
From Eq.~(\ref{eq:telement}) it is obvious that the nondiagonal t- matrix element $t_a(\om)$ is only non-zero in the superconducting state.  It corresponds to an Andreev-type scattering process where holes scatter into electrons and vice versa due to the presence of the condensate.
%
\begin{figure}
\centerline{
\includegraphics[width=0.85\linewidth]{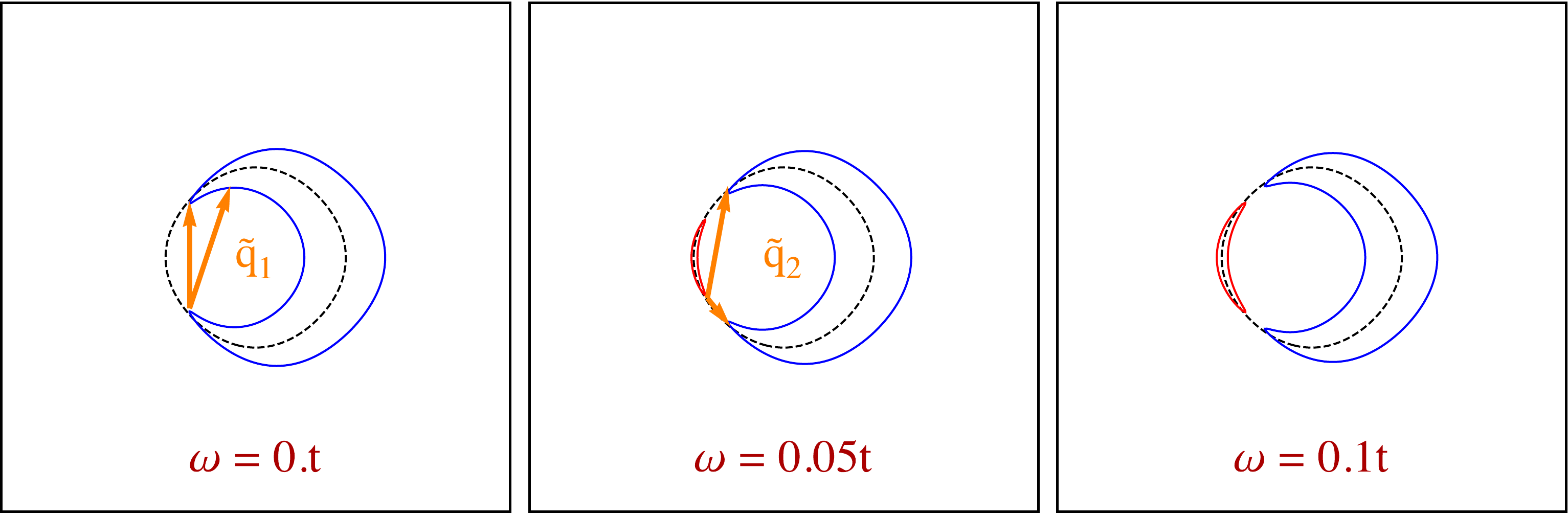}
}
\centerline{
\includegraphics[width=0.85\linewidth]{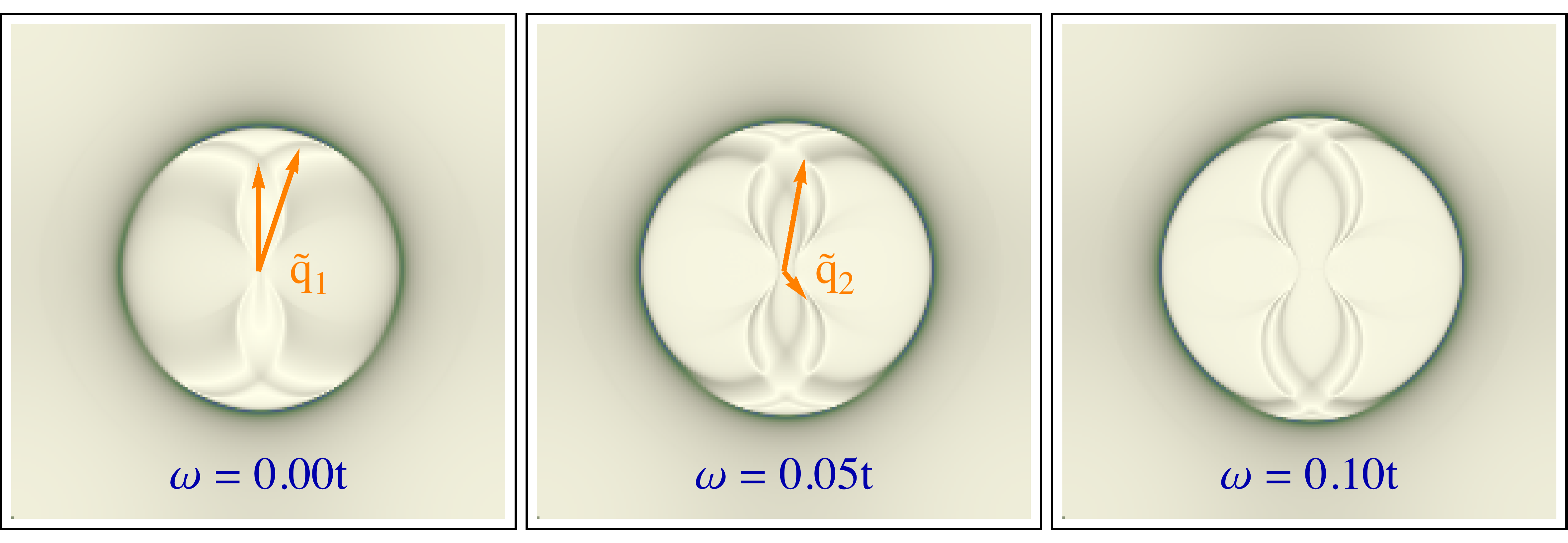}
}
\caption{Spectral function and associated QPI spectrum for $h=0.9\De_0$ with $q(h)= 0.086, \Delta_q(h)=0.115\De_0$.
Top: Spectral function in $(k_x,k_y)$-plane (units: $\pi/a$, range $[-1,1]$). Fermi surface for $\mu=-2.8t$ shown black dashed. We assume $\bq={q}\hbx$. For $\omega=0$ the unpaired region in \bk-space with $E^-_{\bk\bq} <0$ (blue) correspond to regions inside the full line. For finite $\omega$ the evolution of equal energy contours $-E^-_{\bk\bq} =\omega$ (blue) and $E^+_{\bk\bq} =\omega$ (red) is shown.
Bottom: Corresponding QPI spectrum in $(\tq_x,\tq_y)$ - plane in Born approximation according to Eq.~(\ref{eq:qpistruc}) for field as above within FFLO phase. Intra- and inter-sheet scattering vectors $\tbq_1$ and  $\tbq_2$ leading to two-pronged arc structure of QPI image are indicated correspondingly.}
\label{fig:QPI_h09}
\end{figure}
%
Now we may evaluate the QPI spectrum using general Eqs.~(\ref{eq:QPI-FFLO}-\ref{eq:telement}) for the FFLO state. The spectral function of Eq.~(\ref{eq:QPI-FFLO}) has two contributions due to diagonal and off-diagonal (Andreev-type) scattering in Nambu space, i.e.  $\tL_0(\tbq,\om)=\tL^d_0(\tbq,\om)+\tL^a_0(\tbq,\om)$ which are individually given by
\bea
\tL^d_0(\tbq,\om)&=&\frac{1}{N}\sum_\bk\frac
{\tilt_\bq^d(\om)(\om+\xi^\da_{\bk-\bq})(\om+\xi^\da_{\bk-\tbq-\bq})-t^d_\bq(\om)|\De_\bq|^2}
{D_{\bk\bq}(\om)D_{\bk-\tbq,\bq}(\om)},
\non\\
\tL^a_0(\tbq,\om)&=&-\frac{1}{N}\sum_\bk\frac
{t_\bq^a(\om)\De_\bq[(\om+\xi^\da_{\bk-\bq}) + (\om+\xi^\da_{\bk-\tbq-\bq})]}
{D_{\bk\bq}(\om)D_{\bk-\tbq,\bq}(\om)},
\label{eq:QPI_tmat}
\eea
with $D_{\bk\bq}(\om)=(\om-E^+_{\bk \bq})(\om+E^-_{\bk\bq})$.
This result demonstrates directly that the off-diagonal parts are only present in the superconducting state. It is easy to see that a substitution $\bk\leftrightarrow \bk-\tbq$ which leaves the integral invariant leads to the symmetry $\tL^{d,a}_0(-\tbq,\om)=\tL^{d,a}_0(\tbq,\om)$ and therefore the total QPI spectrum also satisfies $\tL_0(-\tbq,\om)=\tL_0(\tbq,\om)$.\\

The simplest way to evaluate this expression is the Born approximation (BA) which consists in taking only single scattering events and no repeated scattering at any given impurity site. Then only terms of first order in the potential strength $V_c$ are considered, meaning  $\tilt^d_\bq(\om)=t^d_\bq(\om)=V_c$ and $t_a(\om)=0$. Therefore the Andreev scattering is absent in Born approximation because it requires at least one intermediate scattering event into the condensate. With the drastic simplification of BA we obtain then for the QPI spectrum $\Lambda_0(\tbq,\om)=V^{-1}_c\tL_0(\tbq,\om)$ of the FFLO state:
\bea
\hspace{-1cm}
\Lambda_0(\tbq,\om)&=&\frac{1}{N}\sum_\bk\frac
{(\om+\xi^\da_{\bk-\bq})(\om+\xi^\da_{\bk-\tbq-\bq})-|\De_\bq|^2}
{(\om-E^+_{\bk \bq})(\om+E^-_{\bk\bq})(\om-E^+_{\bk-\tbq \bq})(\om+E^-_{\bk-\tbq\bq})}.
\label{eq:qpistruc}
\eea
For the conventional BCS case $(h,\bq=0)$ this general-\bq~result simplifies to the well-known expression 
\bea
\Lambda_0(\tbq,\om)&=&\frac{1}{N}\sum_\bk\frac
{(\om+\xi_{\bk})(\om+\xi_{\bk-\tbq})-|\De_\bk|^2}
{[(\om)^2-E^2_\bk][(\om)^2-E^2_{\bk-\tbq}]},
\label{eq:qpistruc0}
\eea
where we now re-introduced a general anisotropic gap function $\De_\bk$ with CM momentum $2\bq=0$. We will use mostly the BA expression of Eq.(\ref{eq:qpistruc}) to investigate the typical momentum-space features in the FFLO state, except when stated otherwise.
%
\begin{figure}
\centerline{
\includegraphics[width=0.9\linewidth]{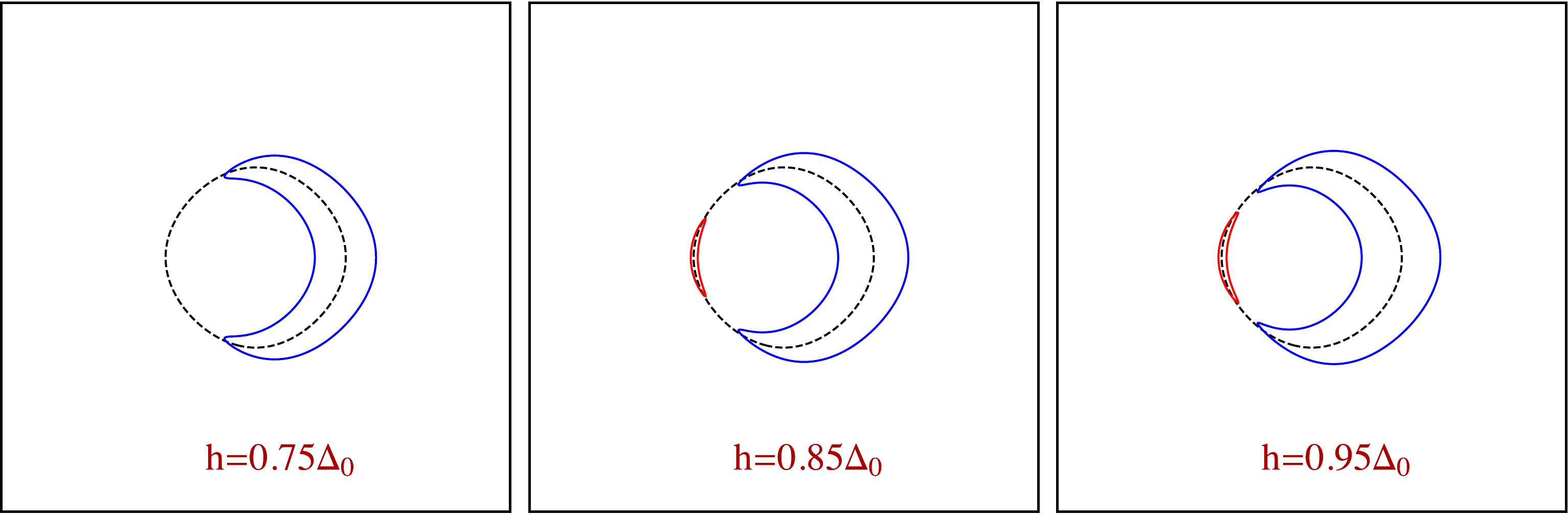}
}
\centerline{
\includegraphics[width=0.9\linewidth]{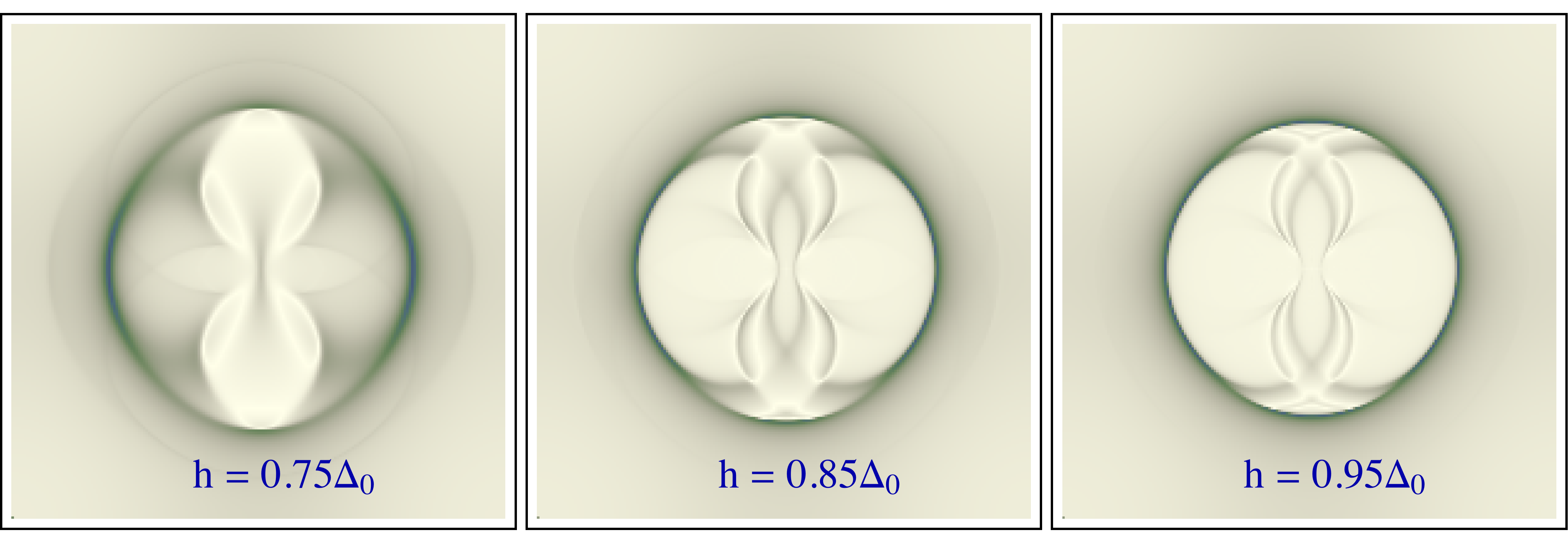}
}
\caption{Spectral function (top) and corresponding QPI image for constant $\omega =0.07t$ as function of field strength $h$ in the FFLO region. For $h<0.7\De_0$ in the BCS phase $\omega\ll\De_0$ the spectral function vanishes and QPI image becomes suddenly extinct. For $h >\De_0$ in the normal phase only the outer $2k_F$ circle remains.}
\label{fig:QPI_om05}
\end{figure}
%

\section{Discussion of QPI momentum space images of the FFLO state}
\label{sec:discussion}

The spectral function of quasiparticles in the FFLO phase exhibits a systematic variation of paired and unpaired regions dimensions in $\bk$- space in Figs.~\ref{fig:QPI_h075},\ref{fig:QPI_h09} (top panels). This is shown for fixed field (within FFLO range) as a function of frequency. The elastic surface scattering events connect the sheets of equal energies. Note however that (in BA) the QPI spectrum is the imaginary part of a convolution product of two Green's functions. Therefore not only on-shell scattering processes contribute because the imaginary part of the product contains the real part of one of the Green's functions. Nevertheless one expects that some features of the spectral function which are characteristic for the breakup into paired and unpaired Fermi surface segments should show up in prominent features of the QPI spectrum as an image in the space of scattering vectors $\tbq$.\\

As a reference starting point we first discuss the BCS $(h,\bq=0)$ case. The spectral function is then given by the two concentric (almost) circular sheets $E_\bk=\omega$ around the Fermi surface. They are directly mapped to two circles in the QPI spectrum with twice the radius as shown Fig.~\ref{fig:QPI_BCS} for $\omega <0$. When the bias voltage approaches $-\De_0$ they coalesce to a single circle of radius $2k_F$. When $|\omega|<\De_0$ inside the gap the QPI image rapidly becomes extinguished.  For positive $\omega$ the picture is qualitatively the same. We note that here and in the following the absolute values $|\Lambda_0(\tbq,\omega)|$ are plotted.\\

Now we discuss the evolution of spectral function and QPI in the FFLO region (Figs.~\ref{fig:QPI_h075},\ref{fig:QPI_h09} ). As noted above for the values $(q(h),\De_q(h))$ that minimize the ground state energy only the blue depaired $\bk_2$ region with  $E^-_{\bk\bq}<0$ can appear since the states for $k_x<0$ always have  $E^+_{\bk\bq}>0$. Therefore the spectral function at $\omega=0$ has only this sheet. For positive frequency the red constant energy quasiparticle sheet   $E^+_{\bk\bq}=\omega$ will however be present and also contribute to QPI spectrum.\\

For a field $h=0.75\De_0$  slightly above $h_p$ the spectral function is shown in Fig.~\ref{fig:QPI_h075} (top). Its equal energy surface $-E^-_{\bk\bq}=\omega$  shrinks (grows) for increasing (decreasing) $\omega$ and at the largest frequency  the (red) $-E^-_{\bk\bq}=\omega$ sheet also appears. The gap $\Delta_q=0.24\Delta_0$ with $q=0.074$ is now much smaller than in BCS case and therefore in the QPI image (bottom)  the basic doubled $2k_F$ FS circles (Fig.~\ref{fig:QPI_BCS})  appear even for moderate $|\omega|$. But most importantly there are {\it additional} features inside the circle which are caused by scattering between states in the unpaired region. A few typical scattering vectors of intrasheet $\tbq_1$- type and intersheet $\tbq_2$- type on the top plot that may be recognized also in the QPI image are indicated. We do not distinguish them individually as they form a continuum. They map out two perpendicular lobe structures oriented along $\tq_x$ and $\tq_y$ - directions that are most prominent for positive $\omega$.  In particular the inter-sheet scattering leads to  two-pronged structures {\it perpendicular} to the CM $\bq$-vector which break the rotational symmetry of the BCS-phase QPI. Their dimensions and orientation therefore contains information on the dimension of FS segmentation in  FFLO phase  and the orientation of $\bq$.\\

This is even more distinct for larger fields $h=0.9\De_0$ where $q= 0.086, \De_q= 0.115\De_0$ a second (red) quasiparticle sheet with  $E^+_{\bk\bq}=\omega$ appears already for smaller positive  $\omega$ in Fig.~\ref{fig:QPI_h09} (top).
  Possible prominent scattering vectors  are again indicated. Those denoted by $\tbq_1$ correspond to intra-sheet  and those denoted by $\tbq_2$ to inter-sheet processes . We stress that most of  these scattering vectors can be scanned over whole sectors of the sheets and form a continuum. Therefore the QPI image should contain open or closed arcs rather than spots which are typical for unconventional nodal BCS superconductors. The corresponeding $\tbq$- space image for $h=0.9\De_0$ is shown in the bottom figure. As before the  predeominant feature in all figures is a $2k_F$ - type scattering circle. When frequency is increased its radius grows slightly but the most typical features for the FFLO state are the distinct arc-like  features appear inside the circle characterised by vectors of $\tbq_2$ -type. Comparison with the corresponding spectral function plot (top) demonstrates that the $\tbq_2$-type scattering corresponds to inter-sheet quasiparticle scattering of the two unpaired regions. These vectors have small length when they are roughly perpendicular to the FFLO vector $\bq=\mbox{q}\hbx$, resulting in the double-pronged arc-like features inside the large circle. The doubling of the arcs reflects the two boundary arcs of the red quasiparticle sheet. When the $\tbq_2$-type vectors are more aligned with the x-axis (with $\bq$) the intersheet scattering contribute only to the major circle like intrasheet $\tbq_1$ processes (Fig.~\ref{fig:QPI_h075}).  The doubling of this major $2k_F$ circle which is still visible in  Fig.~\ref{fig:QPI_h075} (bottom) has now turned into a diffuse halo in Fig.~\ref{fig:QPI_h09} in  because of the much reduced FFLO gap size for  $h=0.9\De_0$.  \\
 
%
\begin{figure}
\centerline{
\includegraphics[width=0.9\linewidth]{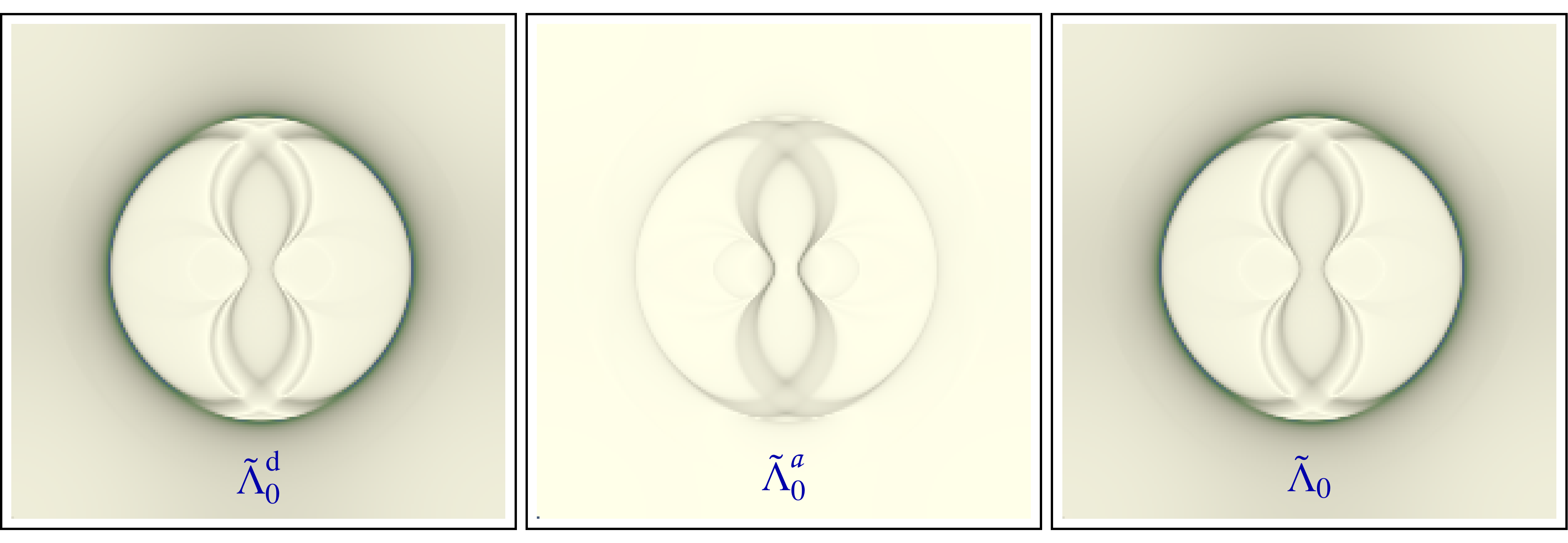}
}
\caption{QPI image calculated with full t-matrix approach (Eq.~\ref{eq:QPI_tmat}) for strong scattering $V_c=5t$ and  $h=0.9\De_0$, $\omega=0.1t$. 
Diagonal ($\tL_0^d$) and anti-diagonal ($\tL_0^a$) t-matrix contributions are plotted separately. The momentum structure of total $\tL_0(\tbq,\omega)$ is quite similar to BA (c.f. Fig~\ref{fig:QPI_h09} (right)) despite the strong scattering limit.} 
\label{fig:QPI_h090_tmat}
\end{figure}
%
In Fig.~\ref{fig:QPI_om05} we show the complementary sequence of spectral function (top) and QPI images for constant bias ($\omega=0.07t$) as function of field in the FFLO field range ($h_p<h<h_c$). Note that for lower fields $h<h_p$ in the BCS phase and $\omega=0$ the spectral function is zero and QPI is extinct. They suddenly appear above $h_p=0.7\De_0$ in the sequence shown in this figure with the typical lobe and arc structures described above. Above $h_c=\De_0$ only the $2k_F$ circle of the normal state remains.\\

Sofar all QPI spectra were discussed within Born approximation. Finally we give an example of the full t-matrix calculation according to Eq.~(\ref{eq:QPI_tmat}) for $h=0.9\De_0$ and $\omega=0.1t$. We use a strong scattering potential $V_c=5t$ to see any differences to the BA result. The t-matrix QPI expression has two contributions coming from diagaonal or anti-diagonal scattering in Nambu space.  They are plotted separately in Fig.~\ref{fig:QPI_h090_tmat} together with the total result. Comparison with BA image in Fig.~\ref{fig:QPI_h09} (right) shows that the overall $\tbq$- momentum structure is well preserved although intensities may be somewhat different. Therefore we conclude that the simple BA already is sufficient to discuss the systematic $\tbq$- momentum structure of QPI in the FFLO phase.
This is because the t-matrix elements in Eq.~(\ref{eq:tmat}) do not have a momentum structure since we assumed isotropic scattering potential $V_c(\tbq)=V_c$ and single-site scattering only. If these conditions are not fulfilled t-matrix and BA results might be quite  different because in the former the frequency and momentum dependence will no longer be nearly factorized as they are in BA.
\\

The analysis given above  has demonstrated that the breakup into paired and two unpaired (in general) regions in the FFLO phase leads to a clear {\it additional} QPI features added to the main $2k_F$ scattering circle that is already present in the s-wave BCS phase at large $\omega >\De_0$. The dimensions of these two-pronged arc-like QPI features change with frequency and field in accordance with the extent of the unpaired regions in the FFLO state. Therefore their mapping in QPI experiments would allow a detailed reconstruction of its partly paired momentum structure.

\section{Conclusion and outlook}
\label{sec:outlook}

In this work we investigated the momentum-space image of the FFLO phase in QPI experiments. To give a proof of principle we have restricted ourselves to a simple one-orbital tight binding Fermi surface and an underlying s-wave gap model. We determined the CM pairing momentum $\bQ=2\bq(h)$ and the associated gap $\Delta_\bq(h)$ of the FFLO state directly by minimizing the ground state energy. Using a Bogoliubov transformation for the case of finite pair momentum the quasiparticle excitation branches and the associated Green's function of the FFLO state were given. The main feature of the FFLO phase, namely the coherent superposition of paired and unpaired regions in \bk-space was discussed by using the properties of the excitation spectrum.

In the main part the latter was used to derive the QPI spectrum within t-matrix theory and the simple Born approximation.  It was shown that the QPI pattern contains additional distinct features not present in the BCS state that are characteristic for the momentum space segmentation of the FFLO state. The geometric dimensions of these QPI-FFLO structures changes continuously with field and frequency. Their orientation is perpendicular to the CM pair momentum. A determination of these image patterns by STM-QPI experiments would allow a detailed mapping of paired and unpaired regions of the FFLO phase in momentum space.

The theory presented here has allowed us to investigate the principal effects that one should expect in FFLO-QPI. It  can be modified in various ways for application to real candidate superconductors. It is straightforward to extend to more complicated Fermi surfaces and non-s wave SC states like $d_{x^2-y^2}$ state realized in the prominent example \CC. There is a chance that 2D multilayer or heterostructure superconductors can exhibit FFLO phase. In this case inversion symmetry breaking at the interface will lead to non-centrosymmetric structure and Rashba terms may appear.  There are investigations how the FFLO state is modified under their presence \cite{loder:13,zhang:13} . Extension of the present QPI theory to this case  appears possible, similar to the BCS non-centrosymmetric superconductors with Rashba spin-orbit coupling\cite{akbari:13,Akbari:2013aa}.\\[0.5cm]


\section*{Acknowledgements}
The authors thank Peter Fulde for helpful discussions. 
They are grateful to the Max Planck Institute for the Physics of Complex Systems (MPI-PKS) for the use of computer facilities.
A.A. wish to acknowledge the Korea Ministry of Education, Science and Technology, Gyeongsangbuk-Do and Pohang City for Independent Junior Research Groups at the Asia Pacific Center for Theoretical Physics. 
The work by  A.A. was supported through  NRF funded by MSIP of Korea (2015R1C1A1A01052411).  
A.A. acknowledges support by  Max Planck POSTECH / KOREA Research Initiative (No. 2011-0031558) programs through NRF funded by MSIP of Korea. 

\section*{References}
\bibliographystyle{NJP}
\bibliography{References}

\end{document}